\documentclass[bibyear]{aa}  
\usepackage{txfonts}
\usepackage[breaklinks, colorlinks, citecolor=blue, urlcolor = blue]{hyperref}
\usepackage{enumitem}

\usepackage{graphicx}

\def\spose#1{\hbox to 0pt{#1\hss}}
\newcommand\lsim{\mathrel{\spose{\lower 3pt\hbox{$\mathchar"218$}}
     \raise 2.0pt\hbox{$\mathchar"13C$}}}
\newcommand\gsim{\mathrel{\spose{\lower 3pt\hbox{$\mathchar"218$}}
     \raise 2.0pt\hbox{$\mathchar"13E$}}}

\def\ltsima{$\; \buildrel < \over \sim \;$}
\def\lsim{\lower.5ex\hbox{\ltsima}}
\def\gtsima{$\; \buildrel > \over \sim \;$}
\def\gsim{\lower.5ex\hbox{\gtsima}}

\def\apj{ApJ}
\def\apjl{ApJL}
\def\apjs{ApJSS}

\def\aap{A\&A}

\usepackage{xcolor}

\usepackage[normalem]{ulem}

\begin{document}

\title{Proton--synchrotron as the radiation mechanism of the prompt emission of GRBs? }

\titlerunning{GRB emission mechanism}
\authorrunning{Ghisellini et al.}
\author{G. Ghisellini\inst{1}\thanks{E--mail: gabriele.ghisellini@inaf.it},
G. Ghirlanda\inst{1}, 
G. Oganesyan\inst{2, 3, 4},
S. Ascenzi\inst{1},
L. Nava\inst{1,5,6, 7}, 
A. Celotti\inst{8, 1, 6, 7},
O.S. Salafia\inst{1},
M.E. Ravasio\inst{1,9},  M. Ronchi\inst{1} 
}
\institute{$^1$  INAF -- Osservatorio Astronomico di Brera, Via Bianchi 46, I--23807 Merate, Italy \\
$^2$ Gran Sasso Science Institute, Viale F. Crispi 7, I-67100, L’Aquila (AQ), Italy\\
$^3$ INFN - Laboratori Nazionali del Gran Sasso, I-67100, L’Aquila (AQ), Italy\\
$^4$ INAF - Osservatorio Astronomico d’Abruzzo, Via M. Maggini snc, I-64100 Teramo, Italy\\
$^5$ INAF – Osservatorio Astronomico di Trieste, via G.B. Tiepolo 11, 34143 Trieste, Italy\\
$^6$ INFN, -- Sezione di Trieste, via Valerio 2, 34127 Trieste, Italy \\
$^7$ IFPU -- Institute for Fundamental Physics of the Universe, via Beirut 2, 34151 Trieste, Italy \\
$^8$ SISSA, via Bonomea 265, 34136 Trieste, Italy \\
$^9$ Università degli Studi di Milano-Bicocca, Dip. di Fisica ``G. Occhialini'', Piazza della Scienza 3, 20126 Milano, Italy\\
}


\abstract{
We discuss the new surprising observational results that indicate quite convincingly
that the prompt emission of Gamma--Ray Bursts (GRBs) is due to synchrotron radiation produced
by a particle distribution that has a low energy cut--off.
The evidence of this is provided by the low energy part of the spectrum of the prompt emission,
that shows the characteristic $F_\nu \propto \nu^{1/3}$ shape followed by
 $F_\nu \propto \nu^{-1/2}$ up to the peak frequency.
This implies that although the emitting particles are in fast cooling, they 
do not cool completely.
This poses a severe challenge to the basic ideas about how and where the 
emission is produced, because the incomplete cooling requires a small value of the
magnetic field, to limit synchrotron cooling, and a large emitting region, to limit
the self--Compton cooling, even considering Klein--Nishina scattering effects.
Some new and fundamental ingredient is required for understanding the GRBs prompt emission.
We propose proton--synchrotron as a promising 
mechanism to solve the incomplete cooling puzzle.
}

\keywords{
gamma--ray burst: general --- radiation mechanisms: non--thermal 
}
\maketitle

\section{Introduction}

The radiation mechanism of the prompt emission of Gamma--Ray Bursts (GRBs)
has been debated since the very first observations.
Its non--thermal appearance and the idea that shocks are responsible for
accelerating particles and enhancing the magnetic field soon led to the
proposal that the synchrotron process should be the dominant radiative mechanism 
(Katz 1994, Rees \& Meszaros 1994, Tavani 1996).

The observed fast variability (down to the millisecond timescales, e.g. 
Walker, Schaefer \& Fenimore 2000) requires the source to be compact, 
therefore with large magnetic and radiation energy densities. 
In these conditions radiative cooling is very efficient, and the corresponding spectrum
is expected to be $F_\nu\propto \nu^{-0.5}$ or softer  
(e.g. Ghisellini \& Celotti 1999).
The observed spectrum is instead much harder (see e.g. Preece et al. 1998a).
When fitted with the Band function (Band et al. 1993),
that is a phenomenological 
model composed by two smoothly connected broken power laws, the average spectrum 
shows a peak in the $\nu F_\nu$ representation, with 
photon spectral slopes $\alpha\sim 1$ below and $\beta\sim 2.3$ above 
the peak frequency $\nu_{\rm peak}$ ($\dot N_{\nu}\propto \nu^{-\alpha}, \nu^{-\beta}$; 
Kaneko et al. 2006, Nava et al. 2011, Goldstein et al. 2012, Gruber et al. 2014, Lien et al. 2016). 
This remains true when considering time resolved spectra (for the brightest bursts,
e.g. Preece et al. 1998b; 
Ghirlanda, Celotti \& Ghisellini 2002; 
Burgess et al. 2014;
Yu et al., 2016).
Rarely, the very hard low energy spectra have been reproduced with a thermal component: 
in a few cases with a pure black body spectrum  
(Ghirlanda, Celotti \& Ghisellini 2004; 
Ghirlanda, Pescalli \& Ghisellini 2013); 
more often with a power law or a Band model with the addition of a black body contribution
(Ryde \& Pe'er 2009;
Ryde et al. 2010; Guiriec et al. 2011; Burgess et al. 2014;
Pe'er \& Ryde 2017; but see Ghirlanda et al. 2007). 

\begin{figure} 
\vskip -0.6 cm
\includegraphics[width=9.6cm]{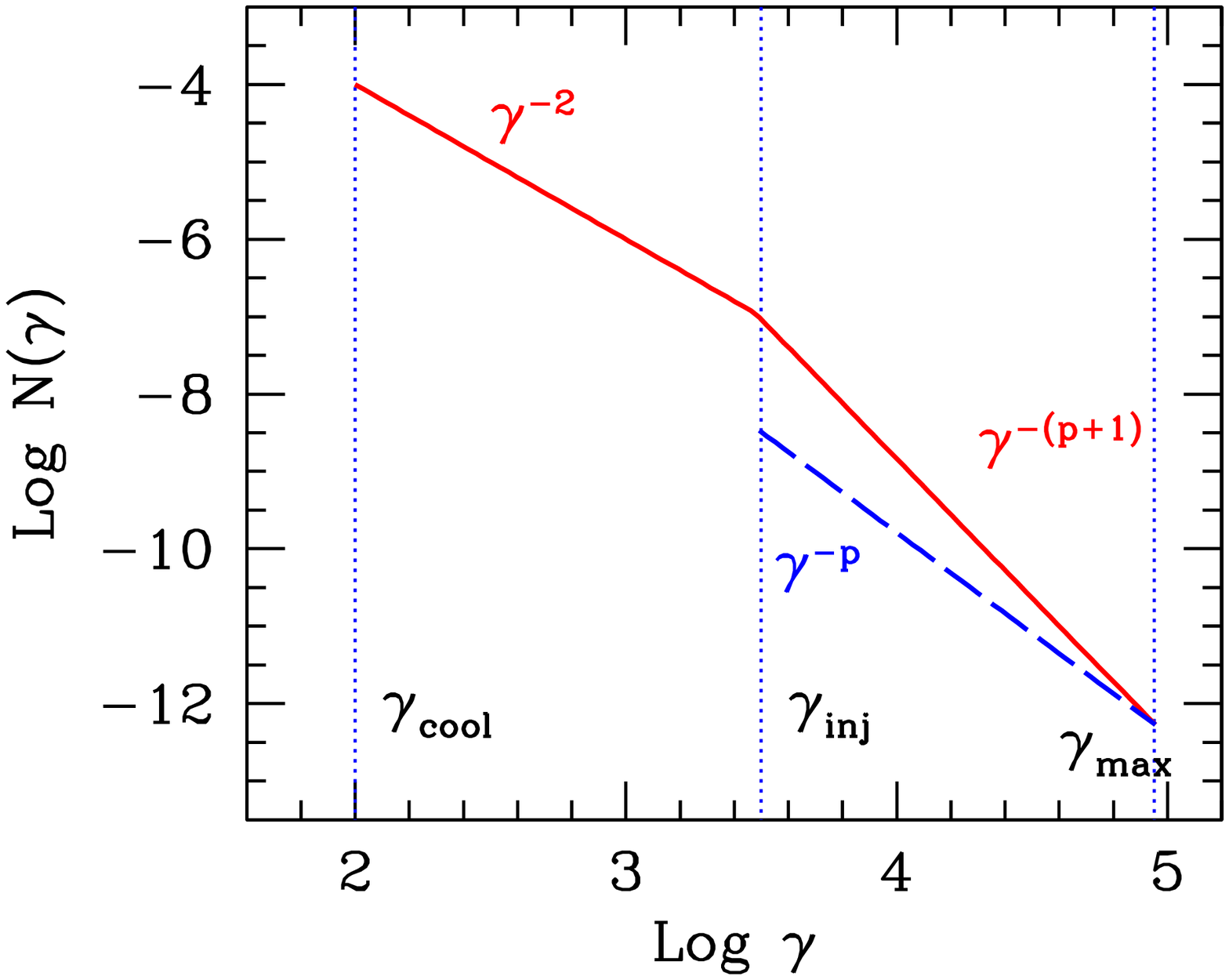}  
\vskip -2 cm
\includegraphics[width=9.8cm]{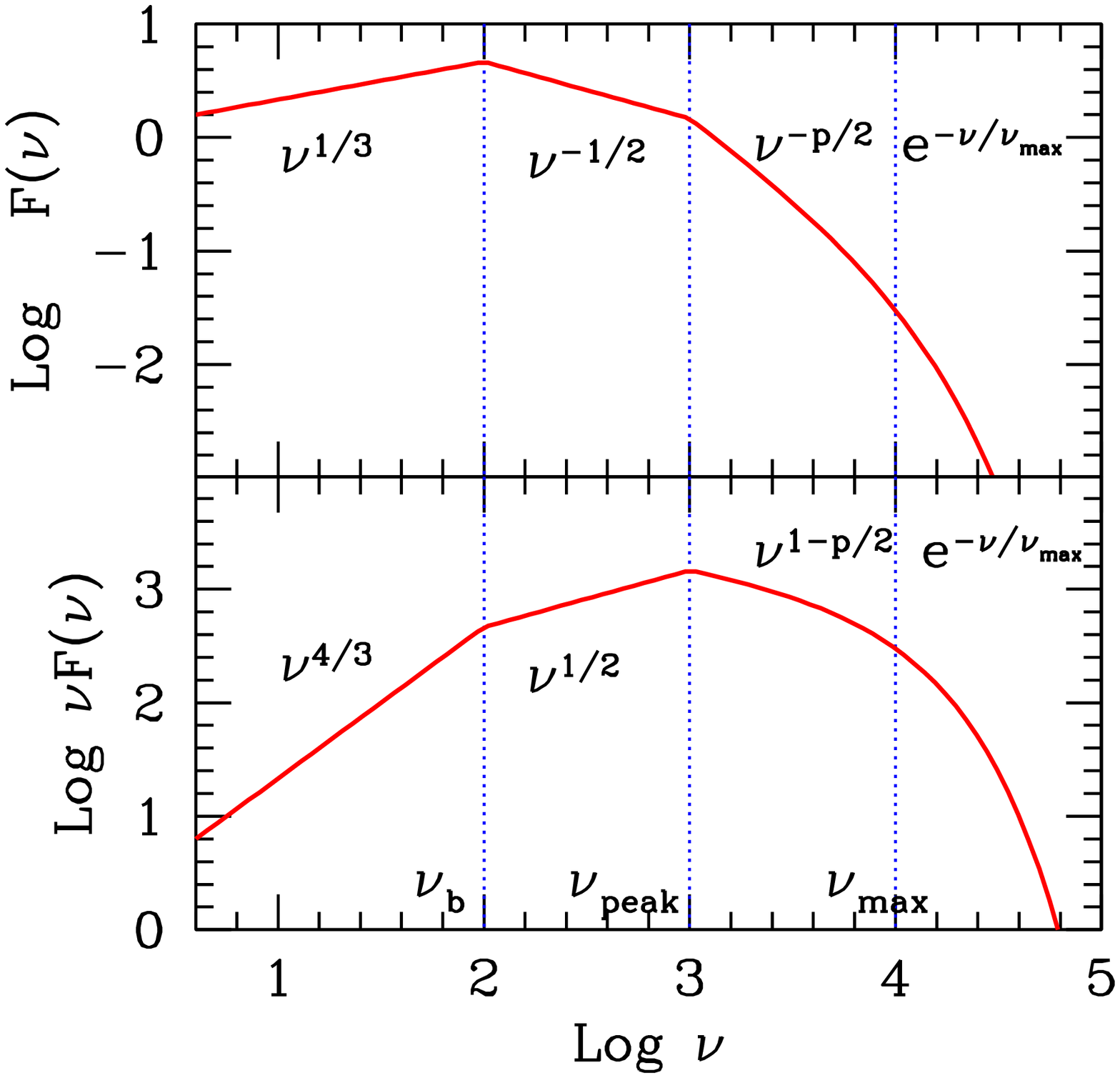}  
\vskip -0.5 cm
\caption{
Top panel: schematics of the particle distribution responsible for the spectra of the bottom two panels.
The dashed blue line corresponds to the injected [$Q(\gamma)$] distribution. 
The characteristic Lorentz factors and frequencies are labeled as described in the text.
The solid and dashed lines are re--scaled by an arbitrary amount.
Bottom two panels: 
sketch of synchrotron spectra as reproduced by the spectral analysis  discussed
in the text. 
The spectra show a high energy exponential cut off which is not always present/detectable in real data.
}
\label{fv}
\end{figure}

Recently, it has been realized that the overall spectral energy distribution (SED) 
could be fitted by three power laws, smoothly joining at two energies: 
one at the break frequency $\nu_{\rm b}$ and the other at the peak frequency $\nu_{\rm peak}$ 
(Oganesyan et al. 2017, Oganesyan et al. 2018, Oganesyan et al. 2019, Ravasio et al. 2018, 
Ravasio et al. 2019). 
Below $\nu_{\rm b}$ the photon spectral index is close to $\alpha_1=2/3$; 
between $\nu_{\rm b}$ and $\nu_{\rm peak}$ the index approximates $\alpha_2=1.5$
and above $\nu_{\rm peak}$, the index $\beta$ becomes (as before - Nava et al. 2011) close to 2.3 
 or slightly steeper ($\beta=$2.8) when allowing for the presence of another break at low energies, possibly with an exponential cut off at high energies. 
This resulting typical spectrum is sketched in the two bottom panels of Fig. \ref{fv}. 

More physically, Oganesyan et al. (2019) also successfully reproduced GRB spectra 
with the synchrotron spectrum 
produced by a non--thermal electron energy distribution 
(see also Chand et al. 2019, Burgess et al. 2019, Ronchi et al. 2019). 
The top panel of Fig. \ref{fv} shows the particle distribution
corresponding to the assumption that it emits such synchrotron radiation.
It must have a low energy cut off at some energy $\gamma_{\rm b}=\gamma_{\rm cool}$ and 
particles close to 
$\gamma_{\rm cool}$ are responsible for the emission with the hard index $\alpha_1$.
The value of the index $\alpha_2$ strongly suggests that the corresponding emitting particles 
are radiatively cooling and distributed as $N(\gamma) \propto \gamma^{-2}$.
Above $\gamma_{\rm peak}=\gamma_{\rm inj}$, $N(\gamma)$ must be a relatively steep power law, 
$N(\gamma)\propto \gamma^{-3.6}$, to account for the observed $\beta=2.3$.

The particle distribution $N(\gamma)$ can be obtained considering particle injection and 
radiative cooling. 
Suppose to inject, throughout an emitting source of size $R$, 
relativistic particles at a rate  
$Q(\gamma) \propto\gamma^{-p}$ between $\gamma_{\rm inj}$ and $\gamma_{\rm max}$, 
as shown by the dashed line in the top panel of Fig. \ref{fv}. 

If the radiative cooling rate is $\propto \gamma^2$, the emitting 
particle distribution $N(\gamma, t)$, after one light crossing time $R/c$ 
[$N(\gamma, R/c)$] is schematically characterized
by the red line in the top panel of Fig. \ref{fv} in the case of fast cooling (i.e. when
$\gamma_{\rm cool}<\gamma_{\rm inj}$).
We have that:
\begin{enumerate}
\item there are no particles below $\gamma_{\rm cool}$ and above $\gamma_{\rm max}$;
\item $N(\gamma)\propto \gamma^{-2}$ between $\gamma_{\rm cool}$ and $\gamma_{\rm inj}$;
\item $N(\gamma)\propto \gamma^{-(p+1)}$ between $\gamma_{\rm inj}$ and $\gamma_{\rm max}$.
\end{enumerate}

Such particle distribution emits a synchrotron spectrum:
\begin{enumerate}
\item $F_\nu \propto \nu^{1/3}$ for $\nu < \nu_{\rm cool}$.
This low energy tail is mainly produced by particles with random Lorentz factor 
$\gamma_{\rm cool}$;

\item $F_\nu\propto \nu^{-1/2}$ between $\nu_{\rm cool}$ and 
$\nu_{\rm peak}$, radiated by particles
with random Lorentz factors  $\gamma_{\rm cool} < \gamma < \gamma_{\rm inj}$;

\item $F_\nu\propto \nu^{-p/2}$ in the range from $\nu_{\rm peak}$  to
$\nu_{\rm max}$, owed to particles with 
$\gamma_{\rm inj} < \gamma < \gamma_{\rm max}$.

\item Above $\nu_{\rm max}$, the spectrum ends with an exponential cut.
The emission is basically emitted  by particles with $\gamma_{\rm max}$. 
\end{enumerate}

For $p> 2$, the $\nu F_\nu$ spectrum peaks at the frequency mainly produced by electrons
with random Lorentz factors $\gamma_{\rm inj}$ (the example shown in Fig. \ref{fv}), while 
for  $p< 2$, the spectral peak corresponds to the frequency chiefly emitted by electrons
with $\gamma_{\rm max}$.

%
%
The fact that the majority of the spectra of  bright long  GRBs can be fitted with 
the above mentioned three-power law model (Oganesyan et al. 2017; Ravasio et al. 2018, 2019) indeed 
suggests that the synchrotron process is 
the radiative mechanism originating the prompt emission.
This implies that the emitting particles {\it do not cool  completely} 
(Daigne, Bosnjak \& Dubus 2011),
but ``remain" at the energy $\gamma_{\rm cool}$ for a timescale
comparable to the typical time bin of the time resolved spectral analysis ($\sim$ 1 s).
This poses a challenge, since the prompt emission is believed to 
be produced in compact regions, as demonstrated by the very rapid variability 
of the flux, that can reach values as short as one millisecond (i.e. Bhat et al. 1992).
Even accounting for the relativistic Doppler time contraction, the emitting region must
be small and located at a distance $R\le ct_{\rm var}\Gamma^2/(1+z)$ from the central engine.
This in turn must correspond to large energy densities, both magnetic and radiative,
leading to very efficient radiative cooling due to the synchrotron and self Compton processes. 
In the above scenario however 
cooling should "stop" when particles reach values of 
$\gamma_{\rm b}=\gamma_{\rm cool}$ significantly larger than unity.

As specified below, in this framework incomplete cooling of the electrons 
would demand low magnetic field (to avoid fast synchrotron cooling), 
and large radii (to avoid fast inverse Compton cooling), but the observed 
short variability timescales require small radii.
This is the key issue we face in this work. 
  
In \S 2 we reassess the synchrotron and self-Compton cooling and their relative relevance. 
Estimates on the expected magnetic field are revised in \S 3. 
We examine ways out within the "standard" scenario in \S 4. 
A proposed alternative, namely proto-synchrotron radiation, is presented in \S 5, 
while we present our conclusions in the final \S 6.   
  
Hereafter we adopt the notation $Q=10^x Q_x$ and cgs units, unless otherwise noted, and 
a flat cosmology with $\Omega_\Lambda=h=70$.

\section{Radiative cooling timescale}
In this section we estimate the cooling timescale of leptons
emitting synchrotron and self--Compton radiation.
In general, the self--Compton process will occur 
partly in the Thomson and partly in the Klein--Nishina regime.
As detailed below, since the latter process is less efficient, it will be approximated.

When treating the inverse Compton (IC)
scattering it is convenient to adopt 
dimensionless photon energies $x\equiv h\nu/(m_{\rm e} c^2)$.
In the comoving (hereafter primed) frame 
the scattering is described by the Klein--Nishina cross section $\sigma_{\rm KN}$ which 
equals the Thomson one ($\sigma_{\rm T}$) for $x^\prime \ll 1/\gamma$.
For simplicity, we then assume that
\begin{eqnarray}
\sigma_{\rm KN} \, &= \, \sigma_{\rm T},  \quad x^\prime \le 1/\gamma \nonumber \\
\sigma_{\rm KN} \, &= \, 0,  \quad \quad x^\prime > 1/\gamma.
\end{eqnarray}
This overestimates somewhat the cross section when $x^\prime \sim 1/\gamma$ (in this case
$\sigma_{\rm KN} =0.43 \sigma_{\rm T}$) and of course underestimates it
at high energies.
However, this approximation is reasonable when considering scatterings between rather
wide distributions of photon and electron energies, becoming more inaccurate when
these are narrow.

According to such approximation, an electron of random Lorentz factor $\gamma$ loses energy 
by scattering a fraction of the total radiation energy density $U^\prime_{\rm r}$, given by
\begin{equation}
U^\prime_{\rm r} = { L^\prime_{\rm iso}\over 4\pi R^2 \Delta R^\prime} {\Delta R^\prime\over c}
= {L_{\rm iso} \over 4\pi R^2 c \Gamma^2}. 
\end{equation}
Not all this radiation energy density is available for scattering in the 
Thomson regime. 
The larger $\gamma$ the smaller the fraction $f(\gamma)$ of scattered photons:
\begin{equation}
f(\gamma)\, = { \int^{1/\gamma}_0 U^\prime_{\rm r}(x^\prime)\, dx^\prime \over U^\prime_{\rm r}}. 
\end{equation}
The corresponding lepton cooling rate can be expressed as:
\begin{equation}
P_{\rm IC}(\gamma)\, = \dot\gamma_{\rm IC} m_{\rm e}c^2\, =\, 
{4\over 3} \sigma_{\rm T} c \gamma^2 U^\prime_{\rm r}\, f(\gamma),
\end{equation}
and it is accurate enough to estimate the cooling time of electrons 
emitting by the synchrotron and IC process, namely:
\begin{equation}
t^\prime_{\rm cool}(\gamma)\, = {\gamma\over \dot\gamma} =
{ 3 m_{\rm e}c^2 \over 4 \sigma_{\rm T} c\gamma \left[ U^\prime_{\rm B}+
U^\prime_{\rm r}\, f(\gamma) \right] },
\end{equation}
where $U^\prime_{\rm B}$ is the magnetic field energy density.
For a source at a redshift $z$ whose flow is moving relativistically, 
the observed cooling timescale appears
$ t^{\rm obs}_{\rm cool} = t^\prime_{\rm cool}(1+z)/\delta$, where
$\delta=[ \Gamma(1-\beta\cos\theta)]^{-1}$ 
is the relativistic Doppler factor and $\theta$ is the viewing angle
of the flow with respect to the line of sight. 
Approximating $\delta \sim \Gamma$, and using
\begin{equation}
\nu^{\rm obs} = { 4\over 3} \, {eB^\prime \over 2\pi m_{\rm e} c} 
\gamma^2  
{\Gamma \over 1+z} \,\,  {\rm i.e.,} \,
\gamma= \left[ {  3 \pi m_{\rm e} c \nu^{\rm obs} \over  2 eB^\prime } \, { (1+z)\over  \Gamma  } \right]^{1/2}
\label{vs}
\end{equation}
we obtain
\begin{eqnarray}
t^{\rm obs}_{\rm cool}(\gamma) &=&
{6\pi m_{\rm e}c^2   \over   \sigma_{\rm T} c B^{\prime 3/2} } 
\left[ {2 e \over   3 \pi m_{\rm e} c \nu^{\rm obs} }\, {1+z \over \Gamma}\right]^{1/2}
{1\over \left[1 + f(\gamma) U^\prime_{\rm r}/U^\prime_{\rm B} \right]} 
\nonumber \\ 
&=& {4.7\times 10^{-8} \, (1+z)^{1/2} \over B_6^{\prime 3/2 } 
\left[{\Gamma_2 \nu_{19}^{\rm obs} } \right]^{1/2}  } \, \times \,
{1\over \left[1+f(\gamma) U^\prime_{\rm r} /U^\prime_{\rm B}\right]}  \, \, \, {\rm s}
\label{tc}
\end{eqnarray}
where the first part of Eq. \ref{tc} is the synchrotron cooling time. 
As reference value the random Lorentz factor of electrons emitting photons 
at frequency $10^{19}$ Hz is $\gamma=163\, [(1+z)\nu_{19}/B^\prime_6\Gamma_2]^{1/2}$.

\begin{figure} 
\vskip -0.6 cm
\includegraphics[width=9.5cm]{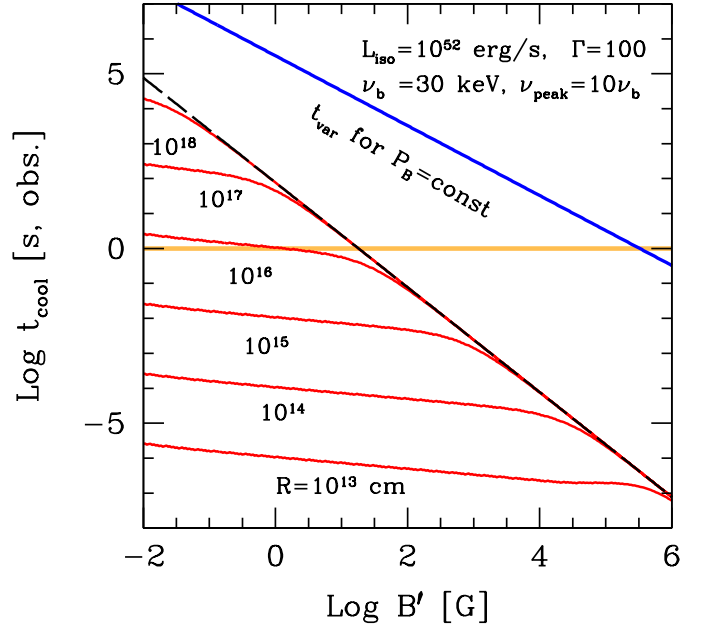}  
\vskip -0.3 cm
\caption{
The observed cooling timescale for the electrons emitting
at the break frequency $\nu_{\rm b}$ as a function of the magnetic 
field and for different distances from the central engine. It is 
assumed that particles cool via synchrotron and self Compton process
(considering, for the latter one, only
the fraction of the synchrotron spectrum below 
$h\nu/(m_{\rm e}c^2) = 1/\gamma_{\rm b}$). 
The black dashed line indicates the synchrotron cooling timescale only.
The blue line is the minimum variability timescale found by assuming that 
the Poynting flux remains constant beyond the acceleration phase, leading
to $B^\prime\propto R^{-1}$ (see \S 3). As reference a redshift 
$z=1$ has been considered.
The orange horizontal line corresponds to a typical exposure time of 1 s.
}
\label{antisyn}
\end{figure}
\begin{figure} 
\vskip -0.6 cm
\includegraphics[width=9.5cm]{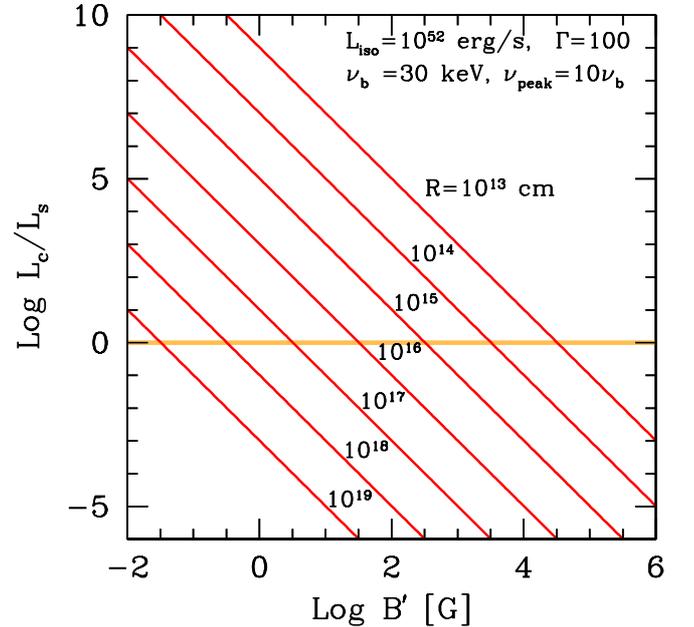}  
\vskip -0.3 cm
\caption{The ratio of the Compton to synchrotron luminosity predicted for the 
same parameters of Fig. \ref{antisyn}.
In this case we have considered the entire distribution of electrons:
for each electron energy, we considered the corresponding synchrotron radiation energy 
that is scattered in the Thomson regime.
The horizontal orange line corresponds to equal synchrotron and self Compton luminosities.
} 
\label{lcls}
\end{figure}
Fig. \ref{antisyn} shows the radiative cooling time for an electron of energy $\gamma_{\rm b}$,
integrating the (comoving) energy density up to the ``Klein Nishina" threshold
$x^\prime_{\rm KN}=1/\gamma_{\rm b}$. 
The radiative cooling time is shown as a function of the magnetic field and for
different distances $R$ from the central engine. 
The black dashed line corresponds to the synchrotron cooling time only. 
The figure shows that, for a given size, a decrease in the magnetic field increases only
slightly the total cooling time because the inverse Compton cooling becomes more severe.
We interpret $\nu_{\rm b}$ as the cooling frequency $\nu_{\rm cool}$.
In Fig. \ref{antisyn} the yellow horizontal line corresponds to one second,
the typical integration time needed to collect enough photons
for the spectral analysis. 
In the case of $t_{\rm cool} \sim 1$ s, the distance $R\gsim 10^{16}$--$10^{17}$ cm and
magnetic fields $B^\prime\lsim 10$ G are required. 

\vskip 0.3 cm

\subsection{Self Compton to synchrotron ratio}

A generic electron of random Lorentz factor $\gamma$
will cool by synchrotron and inverse Compton.
The ratio of the two loss rates is
\begin{equation}
{\dot\gamma_{\rm IC}\over \dot\gamma_{\rm Syn}}
=  { U^\prime_{\rm r} \over U^\prime_{\rm B}} f(\gamma)
\end{equation}
To find the ratio $L_{\rm C}/L_{\rm Syn}$ we must integrate over the particle
energy distribution:
\begin{equation}
{L_{\rm C}\over L_{\rm Syn} } =
{U^\prime_{\rm r} \over U^\prime_{\rm B}}   \,
{ \int_1^{\gamma_{\rm max}} N(\gamma)\gamma^2 f(\gamma) d\gamma   \over 
\int_1^{\gamma_{\rm max}} N(\gamma)\gamma^2 d\gamma   }
\label{eqlcls}
\end{equation}
Therefore we must specify the shape of the particle distribution.
\vskip 0.2 cm

We assume that the typical spectrum observed in the X and $\gamma$--ray energy
range during the prompt emission has the form:
\begin{eqnarray}
F(\nu^\prime) \, &= \,  A \nu^{\prime 1/3},                         
\qquad \qquad\qquad \nu^\prime<\nu^\prime_{\rm b}  \nonumber \\
F(\nu^\prime) \, &= \, A \nu_{\rm b}^{\prime 5/6} \nu^{\prime  -1/2},        
\quad  \nu^\prime_{\rm b}<\nu^\prime<\nu^\prime_{\rm peak}  \nonumber\\
F(\nu^\prime) \, &= \, A \nu_{\rm b}^{\prime 5/6}\nu_{\rm peak}^{\prime \beta-1/2}\nu^{\prime -\beta}, 
\quad \quad   \nu^\prime>\nu^\prime_{\rm peak}.
\label{sed}
\end{eqnarray}
The normalization constant $A$ can be found by the observed total synchrotron flux.
This spectrum is emitted by electrons distributed in energy as a broken power law:
\begin{eqnarray}
N(\gamma) \, &= \, K \gamma^{-2},                 \quad\quad \gamma_{\rm b}\le \gamma\le \gamma_{\rm peak}  \nonumber\\
N(\gamma) \, &= \, K \gamma_{\rm peak}^{p-2}\gamma^{-(p+1)},  \quad \quad   \gamma>\gamma_{\rm peak}
\end{eqnarray}
where $\gamma_{\rm b}$ and $\gamma_{\rm peak}$ are the energies of the electrons
emitting mainly at $\nu_{\rm b}$ and  $\nu_{\rm peak}$.
The slope $p=2\beta$.

Calculating Eq. \ref{eqlcls} assuming the spectrum of Eq. \ref{sed}, we 
constructed Fig. \ref{lcls} showing the Compton to synchrotron luminosity ratio as a
function of the magnetic field and for different distances from the central engine.
It can be seen that to have unimportant Compton emission (i.e. a ratio smaller than unity, represented by orange line)
for magnetic fields $B^\prime<100$ G, the distance $R$ must be larger 
than $\sim 10^{16}$ cm.
This contrasts with the short (sub-second) variability timescales often seen in the
prompt emission of GRBs (see e.g.MacLachlan et al. 2013; McBreen et al. 2001).
In the standard scenario of shells with a spherical curvature, the minimum variability
timescale is, for on axis observers:
\begin{equation}
t_{\rm var} \, =\, {R (1+z)\over 2 c \Gamma^2} \, 
\sim \, 17\, {R_{16}(1+z)\over \Gamma_2^2}\,\, {\rm s}
\end{equation}
This should be compared with the observed variability timescales, that are
much shorter.
To detect short timescales we need a large effective area, and indeed the
fastest (millisecond) variability was detected by BATSE onboard 
the {\it Compton Gamma Ray Observatory} satellite.
Golkhou et al. (2014) reported typical variability timescales of 0.01-1 s with
BAT (Burst Alert Telescope) onboard {\it Swift} and similar variability timescales are observed 
(Golkhou et al. 2015) in the GRBs detected by the Gamma--ray Burst Monitor onboard {\it Fermi}.

\section{Expected magnetic field}

Most models of GRBs require a very large magnetic field at the base of
the jet, to extract the rotational energy of the black hole
through the Blandford \& Znajek (1977) process.
Beyond the acceleration zone of the jet, the Poynting flux $P_{\rm B}$
is assumed to be constant, consistent with the adiabatic assumption. 
The assumption of an initially magnetically dominated fireball is not crucial for our
arguments, and there can be other mechanism able to provide the required energetics
(i.e. neutrino--antineutrino annihilation -- Eichler et al. 1989, Zalamea \& Beloborodov 2011). 
However, it is instructive to derive the value of the magnetic field
in the emitting region, at a distance where the fireball becomes transparent,
in the case of magnetic fields dominating the energetics at the start of the jet.
If all the energy carried by the jet initially (i.e. close to the initial 
radius $R_0$) is magnetic, then the initial $P_{\rm B}$ should be of the same order  
of the total energy $P_{\rm j}$ of the jet after its acceleration. 
The kinetic power is increasing in the acceleration phase at the 
expense of $P_{\rm B}$.
According to this prescriptions, the radial profile of $P_{\rm B}$ 
can be written as:
\begin{equation}
P_{\rm B} = \pi \psi^2 R^2 c \Gamma^2U^\prime_{\rm B} = P_{\rm j} 
\left[1- {\Gamma\beta \over \Gamma_{\rm max}\beta_{\rm max}}(1-\epsilon_{\rm B})  \right]
\end{equation}
where $\epsilon_{\rm B}$ is the fraction of the total power remaining in Pointing flux
after the acceleration phase, and $\psi$ is the semi--aperture angle of the jet.
This leads to a value of the magnetic field, beyond the acceleration zone:
\begin{equation}
B^\prime = \left[ {8 \epsilon_{\rm B} P_{\rm j}  \over c}\right]^{1/2}\, 
{1\over \psi R\Gamma_{\rm max}} 
\label{bfield}
\end{equation}
As an example, for $\epsilon_{\rm B}=0.1$, and $\Gamma_{\rm max}=100$, 
$P_{\rm j}=10^{52}$ erg s$^{-1}$, we have $B^\prime=10^7 \psi^{-1}_{-1}R^{-1}_{13}$ G.
We can relate the value of the magnetic field with the minimum
variability timescale:
\begin{equation}
t_{\rm var} \sim {R\over 2\Gamma^2 c} =  \left[ {8 \epsilon_{\rm B} P_{\rm j}  \over c}\right]^{1/2}\, 
{1\over 2\psi B^\prime\Gamma^3 c} 
\end{equation}
This is the blue line in Fig. \ref{antisyn}: to have short variability timescales, 
the region must be small, and within our approximations (conical jet) this requires
short distances from the central engines, hence large magnetic fields, incompatible
with relatively long cooling timescales.

\section{Ways out}

\subsection{Continuous re-acceleration}

The fast cooling rate could be halted by an acceleration mechanism, dominant at low energies.
If this is constant in time, it means that particles at low energies
are heated, while particles at high energies cool. 
Therefore particles accumulate at the energy for which heating and cooling balance
(see e.g. Asano \& Teresawa 2009; 
see Katz 1994 for the afterglow emission).
There will be a pile up, and the emitted spectrum will disagree with the observed one.
Furthermore, in the internal shock scenario of GRBs, the injected electrons are
always ``new" ones, and they are never reaccelerated.
One could consider a variation of this scheme (Ghisellini, Celotti 1999)
considering a steady state between heating and cooling leading to a thermal 
(e.g. Maxwellian) particle distribution with a sub-relativistic temperature.
In this case the main radiation mechanism is thermal Comptonization,
and the observed spectral indices are unlikely to be obtained (and to be 
the same in different sources), because they require an ad hoc geometrical
and physical set up\footnote{
Thermal Comptonization spectra are usually characterized by a single power law ending 
with an exponential cut, or by a power law, a hump (the ``Wien hump") and 
an exponential tail.
It would be very difficult to obtain the observed three power law segments.
}.

\subsection{Impulsive re-acceleration}

There can be a specific acceleration mechanism that avoids the pile up
of particles.
Assume that the particles are accelerated in a timescale shorter than their
cooling time, and then radiate and cool down to the required $\gamma_{\rm cool}$.
Once reaching $\gamma_{\rm cool}$, they are reaccelerated back to high energies.
This process avoids the pile up of particles.
As an illustrative example, consider some ``acceleration centers"
throughout the source. 
They accelerate particles in a very short timescale.
Immediately after being accelerated, the particles leave the center and 
travel (more or less in random directions) and cool. 
After some time, they arrive to another acceleration center, where they are reaccelerated.
The minimum energy of the particles corresponds to the mean particle travel time from
one acceleration center to another one. 
In our case, since the (comoving) cooling time is of the order of $10^{-5}$ 
seconds (see Eq. \ref{tc})
the average distance among the acceleration centers must be  
$\beta c t_{\rm cool}\sim 3\times 10^5$ cm.
One interesting possibility has been proposed by Sironi, Giannios and Petropoulou (2016)
who studied a scenario of magnetic reconnection in blobs  that are accelerated within the jet by
magnetic tension and can then further accelerate particles.
However, also in this case we again require that the spectrum is produced by the same particles
that are re--used many times.

\subsection{Mini-jets}

We can think to an emitting region that is at large distances 
from the central engine, but is split in many mini-jets, 
and we are observing only one of them
(e.g. Yamazaki, Ioka \& Nakamura 2004, Zhang \& Zhang 2014, Burgess et al. 2019, see also 
Giannios, Uzdensky \& Begelman 2009 for mini jets in blazars).

If the mini-jet is small enough, its emission could vary in
a timescale compatible with what is commonly observed ($10^{-3}-0.1$ seconds).
On the other hand, if this occurs, then the emitting region is compact,
and its radiation energy density would be large, because all
the synchrotron radiation has to be emitted in a small volume.
This implies a dominating Self-Compton emission, that would inevitably 
imply a very fast cooling of particles of all energies.
This problem could be alleviated assuming that the mini-jets are emitting regions
moving with a Lorentz factor $\Gamma_{\rm mini-jet}^\prime=$1--10 as measured in the comoving
frame of the outflow, itself moving with $\Gamma$ with respect to the observer.
This case is equivalent to mini-jets moving with a total 
$\Gamma_{\rm mini-jet}\sim \Gamma\Gamma_{\rm mini-jet}^\prime$
(see e.g. Burgess et al. 2019).
If mini--jets occupy a large fraction of the emitting volume, than there is no difference
with the case of a unique jet moving at $\Gamma_{\rm mini-jet}$.
If instead they occupy a small fraction of the available volume, they could explain fast 
variability, but the carried energy would be a small fraction of the total.
Therefore the efficiency (i.e. the ratio between the radiated and the total jet kinetic 
energy) would be smaller than usually thought.

\subsection{Break due to Inverse Compton in Klein--Nishina regime}

The hard low energy spectrum of GRB emission could be due to the effects of inverse Compton
scatterings occurring in the Klein--Nishina regime (Rees 1967), as suggested by 
Derishev, Kocharovsky \& Kocharovsky (2001); Nakar, Ando, \& Sari (2009) 
Daigne, Bosnjak \& Dubus (2011).
These models are based on the idea that the inverse Compton process is dominant in cooling
the intermediate energy electrons, responsible for the low energy X--rays before $\nu_{\rm peak}$.
Electrons in this energy range cool at a rate $\gamma^a$, with $a<2$, hardening their
energy distribution with respect to scatterings in the Thomson regime.
Electrons at higher energies, responsible for the emission above $\nu_{\rm peak}$,
cool only by synchrotron at a rate $\propto \gamma^2$. 

These models work in a limited range of physical parameters,
since the inverse Compton cooling is required to be reduced, but nevertheless important,
before $\nu_{\rm peak}$, and negligible above. 
Even when this constrain is satisfied, the typical obtained spectral indices are $F_\nu\propto \nu^0$ (see e.g. Fig. 2 of Daigne, Bosnjak \& Dubus 2011).
Harder spectra can be obtained if the adiabatic cooling is important,
and they approach $F_\nu\propto \nu^{1/3}$ very rarely.
(see e.g. Fig. 4 of Daigne, Bosnjak \& Dubus 2011).
In general, in this model, the inverse Compton flux must be important, 
while it is instead limited by the existing observations.

\section{Proton--synchrotron}

A possible solution to the problem of incomplete cooling of the emitting particles
is to assume that what we see is synchrotron radiation produced
by protons, not by leptons (see Gupta \& Zhang 2007 for a discussion of lepton and 
hadronic models for the high energy prompt emission in GRBs and 
Aharonian 2002 for the a proton--synchrotron model applied to blazars).
Protons are accelerated efficiently in shocks, and should receive most of the
shock energy, more than the leptons.
The typical synchrotron frequency emitted by protons is, in the comoving frame:
\begin{equation}
\nu^\prime_{\rm S, p} = {4\over 3} { e B^\prime \over 2 \pi \, m_{\rm p} c} \, \gamma^2
\to \gamma = \left[ 3 \pi\, \nu^\prime_{\rm S, p}  m_{\rm p} c \over 2 eB^\prime \right]^{1/2}
\sim 10^4 \, \left[\nu^\prime_{\rm S, p, keV} \over B^\prime_6 \right]^{1/2}
\label{vsp}
\end{equation}
The total power emitted, for a tangled magnetic field and an isotropic distribution of pitch angles is:
\begin{equation}
P_{\rm S, p} = {4\over 3} \sigma_{\rm T} c  \left(  { m_{\rm e}  \over m_{\rm p} } \right)^2  
{ B^{\prime\, 2} \over 8\pi} \gamma^2
\end{equation}
The synchrotron cooling time (in the observer frame) is:
\begin{eqnarray}
t^{\rm obs}_{\rm cool, S, p} &=&  { \gamma m_{\rm p} c^2  \over P_{\rm S, p} } = 
{ 6 \pi m_{\rm p} c^2   \over   \sigma_{\rm T}c   B^{\prime\, 2}  \gamma } \, 
\left(  { m_{\rm p}  \over m_{\rm e} } \right)^2 {1+z \over \Gamma} \\ \nonumber
&=& 
{ 6 \pi m_{\rm e}c^2     \over   \sigma_{\rm T}c   B^{\prime\, 3/2} } \, 
\left(  { m_{\rm p}  \over m_{\rm e} } \right)^{5/2} 
 \left[ {2 e \over  3 \pi\, c\nu^{\rm obs}_{\rm S, e} } \, {1+z \over \Gamma} \right]^{1/2} \\ \nonumber
&=&
t^{\rm obs}_{\rm cool, S, e}  \left(  { m_{\rm p}  \over m_{\rm e} } \right)^{5/2} 
\sim 1.44\times 10^8 t_{\rm cool, S, e}   \,\,\, {\rm for \, the \, same\, \nu^{\rm obs}_{\rm S} }
\end{eqnarray}

Comparing with the electron synchrotron cooling timescale of Eq. \ref{tc}
we have values close to one second, as observed.

Having an observed cooling timescale of approximately one second
for particles emitting at the observed frequency of 100 keV then requires:
\begin{itemize}
\item[A)] emitting electrons:
a weakly magnetized ($B^\prime\sim$ 1 G, to avoid extremely fast synchrotron cooling) 
and very large ($R\gsim 10^{16}$ cm, to avoid too fast self Compton cooling)  
emitting region;
or
\item[B)] emitting protons: 
a standard magnetic field and emitting region size, namely
$B^\prime\sim 10^6$ G and $R\sim 10^{13}$ cm. 

\end{itemize}

\subsection{Maximum frequency in proton--synchrotron}

Guilbert, Fabian \& Rees (1983) suggested that for shock accelerated electrons, there
is a maximum synchrotron frequency that can be emitted, independent of the random Lorentz factor
$\gamma$ and the magnetic field $B$.
The argument was that each shock crossing, the electrons double its energy, until its
gyro-radius becomes so large that synchrotron cooling limits the maximum attainable $\gamma$.

We can repeat the original argument for protons\footnote{In this subsection all quantities are considered in the comoving frame.}.
We have that ($\beta =\sin\theta\sim 1$, where $\theta$ is the pitch angle):
\begin{equation}
{\Delta \gamma \over \gamma \Delta t} = {1 \over \Delta t}
\to \dot \gamma_{\rm h} = {\gamma \over 2\pi  r_{\rm L}/c } = {eB \over 2\pi m c   } 
\label{gdoth}
\end{equation}
where we set $\Delta t = 2 \pi r_{\rm L}/c$, and   $r_{\rm L}= \gamma m c^2   /(e B)$ 
is the Larmor radius.
The synchrotron cooling rate is: 
\begin{equation}
\dot \gamma_{\rm cool, S} =   {2\over 3} {e^4 \over   m^3 c^5}\gamma^2 B^2 
\label{gdotc}
\end{equation}
Equating Eq. \ref{gdoth} with Eq. \ref{gdotc} we have: 

\begin{equation}
{  eB  \over  2 \pi   m    c  } =  {2\over 3} {e^4 \over   m^3 c^5}  \gamma^2 B^2  
\to 
\left[ \gamma^2 B\right]_{\rm max} = { 3\,   \over 4\pi } {m^2 c^4\over e^3  }
\end{equation}
Therefore, the maximum synchrotron frequency is:
\begin{eqnarray}
h\nu_{\rm s,max} &=&{4\over 3} {he  \over 2\pi m c} \left[ \gamma^2 B\right]_{\rm max} =
{1  \over 2 \pi^2 } {h m  c^3 \over e^2    }   \nonumber  \\
&=& 22 \,\,\, {\rm MeV}, \quad {\rm for \, electrons}\nonumber  \\
&=& 41 \,\,\, {\rm GeV}, \quad {\rm for \, protons}
\end{eqnarray}

\subsection{Total energy and number of emitting particles}

In the standard scenario, the emitting particles are accelerated at the shocks and cool,
and are not re--accelerated. 
Therefore the total number of particles $\cal{N_{\rm iso}}$ contributing to the observed emission is:
\begin{equation}
{\cal N_{\rm iso}} \, \sim \, {E_{\rm iso} \over 
\Gamma mc^2 (\gamma_{\rm inj} -\gamma_{\rm cool}) }
\label{niso}
\end{equation}
This assumes that the slope of the injected distribution is $p>2$. 
We now compare case A (electrons) and case B (protons) assuming in any case
$\Gamma=10^2 \Gamma_2$.

\vskip 0.2  cm

{\it Case A: electrons} --- 
From Eq. \ref{vs} the typical Lorentz factor $\gamma_{\rm cool}$
of the electrons emitting at $\nu_{\rm cool}$  is
\begin{equation}
\gamma_{\rm cool} =  2.5\times 10^4 
\left[ {\nu^{\rm obs}_{\rm cool, keV} \over B^\prime } \, { (1+z) \over \Gamma_2  } \right]^{1/2}
\end{equation}
This leads to a total number of emitting electrons:
\begin{equation}
{\cal N_{\rm e, iso}} \, \sim \, 4.9\times 10^{52} 
{ E_{\rm iso, 53} 
 \over   (\gamma_{\rm inj}/\gamma_{\rm cool}-1)  } 
 \left[ {B^\prime \over\nu^{\rm obs}_{\rm cool, keV}\Gamma_2 (1+z) } \right]^{1/2}
\end{equation}
Observationally, the break $\nu_{\rm b}$, interpreted as the cooling break $\nu_{\rm cool}$, 
is a factor $\sim$ 10 smaller than $\nu_{\rm peak}$. 
This corresponds to $\gamma_{\rm inj}/\gamma_{\rm cool}\sim 3$.

The ratio between the total kinetic energy $E_{\rm K, iso, before}$
(calculated before the prompt emission) and the radiated energy $E_{\rm iso}$ is:
\begin{equation}
{E_{\rm K, iso, before} \over E_{\rm iso} }= 
{(\gamma_{\rm inj}  +m_{\rm p}/m_{\rm e}) \over \gamma_{\rm inj} -\gamma_{\rm cool} }
\end{equation}
This assumes that there is one cold proton per emitting electron. 
The same ratio after the prompt emission is:
\begin{equation}
{E_{\rm K, iso, after} \over E_{\rm iso} }= 
{(\gamma_{\rm cool}  +m_{\rm p}/m_{\rm e}) \over \gamma_{\rm inj} -\gamma_{\rm cool} }
\end{equation}

\vskip 0.2  cm

{\it Case B: protons} --- In this case we assume $B^\prime=10^6 B^\prime_6$ G.
From Eq. \ref{vsp} we have that protons emitting at 1 keV have $\gamma\sim 10^4$.
From Eq. \ref{niso} the total number of emitting protons producing $E_{\rm iso}$ is: 
\begin{equation}
{\cal N_{\rm p,iso}} \, \sim \, 6.9\times 10^{49}
{E_{\rm iso, 53} \over 
(\gamma_{\rm inj}/\gamma_{\rm cool}-1) }
 \left[ {B^\prime_6\over \nu^{\rm obs}_{\rm cool, keV} \Gamma_2 (1+z) } \right]^{1/2}
\end{equation}
In terms of total mass, this corresponds to only $M = N_{\rm p, iso} m_p
\sim 5.5\times 10^{-8} M_\odot$.

Also in this case we can calculate the ratio between the total kinetic energy 
(before the prompt emission) and the radiated energy $E_{\rm iso}$.
Assuming that the leptonic component is unimportant we have:
\begin{equation}
{E_{\rm K, iso, before} \over E_{\rm iso} }= 
{ 1   \over 1- \gamma_{\rm cool}/\gamma_{\rm inj}  } \, \sim \, {3\over 2}
\end{equation}
The same ratio after the prompt emission is:
\begin{equation}
{E_{\rm K, iso, after} \over E_{\rm iso} }= 
{1 \over \gamma_{\rm inj}/\gamma_{\rm cool}-1 }\, \sim \, {1\over 2}
\end{equation}
This indicates that the maximum energy emitted by the afterglow is 
$\sim$ 1/2 of the energetics of the prompt.
This implicitly assumes that the Poynting flux is not the dominant form
of power that can be converted into radiation. 
In the opposite case, we should include the magnetic energy when
calculating  the fraction of the total jet power that can be converted into radiation, 
both in the prompt and the afterglow phases.

Since the emitting protons have $\gamma\gsim 10^4$, greater than $\Gamma$,
is not possible that they derive their energy from the conversion of bulk kinetic
energy into random energy, unless only a minority of protons are accelerated at 
the expense of a much larger population of cold protons. 
This requires a not yet specified mechanism, able to channel a fraction of
the total bulk kinetic energy into a few selected protons. 

Another more likely possibility could be a partial magnetic reconnection of
a dominant magnetic field.
In this case we would have a magnetic dominated flow, with a small baryon loading, and we 
would expect three possible observational
consequences.
The first is the absence of a thermal prompt emission, the ``fossil" radiation 
remaining after the conversion of the internal energy into bulk motion
(see, e.g. Daigne \& Mochkovitch 1998).
The second is polarization of the prompt emission, if part of the magnetic field,
besides being dominant, is also ordered (see e.g. Lyutikov, Pariev \& Blandford 2003). 
The third consequence is a weak or absent reverse shock when the flow starts to 
decelerate (see e.g. Nakar \& Piran 2004)

\subsection{Electron--synchrotron vs proton--synchrotron}

Just for illustration, consider the case in which the number
of injected electrons and protons is the same.
Consider also that the observed spectrum is due to the 
proton--synchrotron process. 
We then ask if the emission produced by electrons can contribute
to the observed prompt flux.

Consider two cases: 
\begin{enumerate}
\item Electrons and protons are injected with the same random
Lorenz factor distribution.
\item Electrons and protons are injected with the same energy distribution.
\end{enumerate}

In case 1), the total  injected power associated to the electron would be a factor
$m_{\rm p}/m_{\rm e}$ smaller, making the electron--synchrotron luminosity
negligible with respect to the proton--synchrotron one.
Furthermore, the  typical frequencies emitted by electron--synchrotron 
would be larger by the factor $m_{\rm p}/m_{\rm e}$ with respect to the 
proton--synchrotron case.

In case 2), if a similar amount of electrons and protons are injected with the
same typical energies, then also the two kinds of bolometric luminosities would be equal,
but the random Lorentz factors of the electrons would be $m_{\rm p}/m_{\rm e}$ times larger.
The typical electron--synchrotron frequencies would be a factor $(m_{\rm p}/m_{\rm e})^3$
larger. 
We are here assuming that the argument leading to a maximum synchrotron emitted
frequency does not apply, requiring an acceleration mechanism different from shocks.
In this case it is likely that this extremely high energy emission ($\sim 10^3$ TeV)
would produce a pair cascade, partly inside the emitting region, and party outside. 
The fraction of luminosity absorbed within the emitting region would reprocess
the power to smaller energies, but a detailed calculation is needed to quantify
this statement. 
The fraction of high energy photons that escape the source can pair--produce 
in the intergalactic medium interacting with the cosmic background light. 
In this case the luminosity, initially collimated into the jet angle,
is dispersed, since the produced pairs would be de--collimated by the 
intergalactic magnetic field. 
It is then likely that the reprocessed light would not contribute to the
observed flux.

\subsection{Radiative cooling and adiabatic timescales}

The proton-synchrotron scenario can work because the radiative cooling timescale for protons
is much longer than for leptons, and this can imprint a signature in the spectrum
(the cooling break at $\nu_{\rm C}$).
On the other hand, one can wonder how we can have a very fast variability (tens of milliseconds)
in this scenario.
The answer lies in the adiabatic timescale, $t_{\rm ad}\sim R/(\Gamma^2 c)$
that is of the same order of the minimum variability timescale.
After $t_{\rm ad}$, the size of the emitting region roughly doubles, 
all particle energies halve, the normalization of the particle distribution
decreases (to conserve the number of emitting particles), as well as the magnetic field.
As a result the emitting flux, even if the radiative cooling is not particularly
severe, is bound to decrease.
The $\nu_{\rm C}$ break continues to evolve (becoming smaller) but the flux decreases, making
this evolution difficult to observe.
In addition, when using a relatively long exposure timescale, we can see the superposition of 
several events, each lasting for $t_{\rm ad}$. 
If all these events have a similar $\nu_{\rm C}$ we will observe a non--evolving 
break frequency (as in the case of GRB 160625B discussed in Ravasio et al. 2018).
Instead, if the flux is produced by a unique shell, spectra taken at different 
times should show an evolving $\nu_{\rm C}$, decreasing in time at least as $t^{-2}$
(or faster, if the magnetic field is decreasing as well).
This should be best visible during the decay phase of a pulse.
We plan to investigate this interesting issue in a future study.

\section{Conclusions}

We have shown that the recent observations of a low energy break in the prompt
spectrum of GRBs, accompanied by the observations of the slopes below and above
the break, strongly suggests that the emission process is synchrotron done by particles
that cannot completely cool.
This is at odds with our expectations about the properties of the emitting regions,
that should be compact and then strongly magnetized. 
We have shown that the size of the emitting region should be quite large, to avoid
a strong self Compton emission (and thus a severe cooling). 
Furthermore, the inferred lower limits on the size can dangerously start to 
conflict with the limits posed by the onset of the afterglow.

In a leptonic scenario we found no simple solution to this problem.
We consider this so serious to need some explanation alternative to the
common and standard scenario we considered up to now (i.e. emitting 
region located just beyond the transparency radius, with strong magnetic field,
very small cooling times, and limited importance of the self--Compton emission).

One possibility able to preserve the standard scenario is to assume that the
radiation we observe is still synchrotron but produced by protons.
Since their random energy exceeds the bulk one, this possibility 
likely requires that the dominant form of energy carried by the jet is
magnetic.
If the magnetic field is also ordered, then we expect a largely polarized
prompt emission. 
A magnetically dominated jet should also imply a limited importance
of any thermal component in the prompt emission, as well as a weak 
(or null) reverse shock at the start of the deceleration phase. 
The first simple estimates concerning the
presence of emitting ultra--relativistic protons are very promising, 
and we plan to further investigate their consequences in the near future.

\section*{Acknowledgements}
We would like to thank Fabrizio Tavecchio for discussion and
the anonymous referee for comments.
We thank a ASI--{\it NuSTAR} grant and  
we acknowledge financial contribution from the agreement ASI-INAF n.2017-14-H.0,
and from the PRIN--INAF 2016 and the PRIN-MIUR ``FIGARO" grants.


\end{document}